\documentclass{article}
\usepackage[margin=1in]{geometry}
\usepackage{amsmath,amsthm,amssymb}
  
\usepackage{algorithm}
\usepackage{algpseudocode}
\usepackage[table]{xcolor}

\newtheorem{construction}{Construction}
\newtheorem{claim}{Claim}
\newtheorem{theorem}{Theorem}
\newtheorem{lemma}{Lemma}
\newtheorem{corollary}{Corollary}
\theoremstyle{definition}
\newtheorem{definition}{Definition}

\title{Improved and Partially-Tight Lower Bounds for Message-Passing Implementations of Multiplicity Queues}
\author{Anh Tran, Edward Talmage\\Bucknell University, Lewisburg, PA, USA}

\begin{document}
\maketitle

\begin{abstract}
  A \emph{multiplicity queue} is a concurrently-defined data type which relaxes the conditions of a linearizable FIFO queue to allow concurrent $Dequeue$ instances to return the same value.  It would seem that this should allow faster implementations, as processes should not need to wait as long to learn about concurrent operations at remote processes and previous work has shown that multiplicity queues are computationally less complex than the unrelaxed version.  Intriguingly, recent work has shown that there is, in fact, not much speedup possible versus an unrelaxed queue implementation.   Seeking to understand this difference between intuition and real behavior, we extend that work, increasing the lower bound for uniform algorithms.  Further, we outline a path forward toward building proofs for even higher lower bounds, allowing us to hypothesize that the worst-case time to $Dequeue$ approaches maximum message delay, which is similar to the time required for an unrelaxed $Dequeue$.  We also give an upper bound for a special case to show that our bounds are tight at that point.  To achieve our lower bounds, we use extended shifting arguments, which have been rarely used but allow larger lower bounds than traditional shifting arguments.  We use these in series of inductive indistinguishability proofs which allow us to extend our proofs beyond the usual limitations of shifting arguments.  This proof structure is an interesting contribution independently of the main result, as developing new lower bound proof techniques may have many uses in future work.
\end{abstract}

\section{Introduction}

In the search for efficient structured access to shared data, \emph{relaxed data types} \cite{HenzingerKirschPayerSezginSokolova13} have risen as an efficient way to trade off some of the precise guarantees of an ordered data type for more performance \cite{TalmageWelch14}.  Multiplicity queues are a recently-developed relaxation of queues \cite{CastanedaRajsbaumRaynal20} which allow concurrent $Dequeue$ instances to return the same value.  Since they cannot have a sequential specification (being defined in terms of concurrency), previous results on relaxed queues do not apply to multiplicity queues.

Multiplicity queues are particularly interesting due to the implications for the computational power of the type.  In \cite{CastanedaRajsbaumRaynal20}, Casta\~{n}eda et al. implement multiplicity queues from $Read$/$Write$ registers, which is impossible for FIFO queues.  This means that it is possible to have queue-like semantics without the cost of strong primitive operations like $Read$-$Modify$-$Write$.  Further work has showed that this allows interesting application in work-stealing \cite{CastanedaPina21}.  In addition, multiplicity queues are more efficient to implement in a shared memory system than the best known algorithm for FIFO queues \cite{JohnenKhattabiMilani22}.

We are interested in message-passing implementations of data types, which provide the simplicity and well-defined semantics of a shared memory system in the message passing model inherent to geographically distributed systems \cite{AttiyaBar-NoyDolev95}.  In queue implementations, the need for concurrent $Dequeue$ instances to wait long enough to learn about each other, so that they can be sure to return different values, is one of the primary reasons that $Dequeue$ is expensive to implement, in terms of time delay from operation invocation to response \cite{WangTalmageLeeWelch18}.   Between the higher performance multiplicity queues achieve in shared memory models and the intuitive notion that allowing concurrent $Dequeue$ instances need not learn about each other to return different values, it seems intuitive that multiplicity queues should be very efficient to implement in a message-passing system.

To the contrary, recent work \cite{Talmage22} showed that there are limited performance gains possible.  In a partially-synchronous system with maximum message delay $d$ and delay uncertainty $u$, that previous result that the worst-case delay from invocation to return for $Dequeue$ of at least $\min\left\{\frac{2d}{3},\frac{d+u}{2}\right\}$, shows that there is at most a factor of 2 speedup, since we know we can implement an unrelaxed queue where $Dequeue$ always returns after $d + (1-1/n)u$ time in this model \cite{WangTalmageLeeWelch18}.

We here extend the work in \cite{Talmage22}, improving the lower bound for the return time of $Dequeue$ in uniform algorithms (those whose behavior does not depend on the number of participating processes) to $\min\left\{\frac{3d+2u}{5}, \frac{d}{2}+u\right\}$.  The intuition is that while a particular $Dequeue$ instance may not need to know about another, concurrent instance, determining which instances are concurrent is expensive in its own right.  This improvement suggests new insights into fundamental properties of message passing implementations of shared data structures, showing that differentiating previous from concurrent instances is the primary driver of the time required for operation instances to choose the correct return value, not trying to detect all concurrent actions.  This could help develop more efficient algorithms or precise relaxations which can find minimal weakening while providing maximum performance improvements.

Except for the edge case of $u=0$, where they match and which we show is tight, our new bound is larger than the previous state of the art, and further gives us better tools to perhaps continue improving the bound.  The proofs in \cite{Talmage22} relied on shifting and other indistinguishability arguments among three or fewer processes, and the bounds were limited by the number of processes.  In this paper, we develop more complex indistinguishability arguments, using inductive definitions of different runs of the algorithm with all processes participating.  This requires using more advanced shifting and indistinguishability tools, similar to those developed in \cite{WangTalmageLeeWelch18}.  These stronger tools allow us to prove larger bounds, and are interesting in their own right as a hint to how we may be able to prove larger lower bounds on other problems, as well.

The piecewise nature of our bound also provides potential insight into what the optimal lower bound may be.  We show that the $\frac{d}{2} + u$ portion of the bound, which seems weaker, is actually tight in the special case when all messages take exactly the same amount of time ($u = 0$).  However, for larger uncertainties in message delay ($u > d/6$), the $\frac{3d+2u}{5}$ portion of the bound is higher.  This means that the lower bound is not even linear in $u$.  In fact, if one can strengthen the base case of our induction, it appears that larger bounds than $\frac{3d+2u}{5}$ may well be possible with the same argument structure, for larger values of $u$, giving some type of curve with slope increasing with $u$.  The base case is already the most complex portion of the proof, so such strengthening and finding an optimal lower bound remain as future work.

\subsection{Related Work}
The idea of relaxing data types grew out of the study of consistency conditions weaker than linearizability.  Afek et al. proposed Quasi-Linearizability in \cite{AfekKorlandYanovsky10}, which requires linearizations to be within a certain distance of a legal sequence, instead of themselves legal.  From another perspective, this is just an expansion of the set of legal sequences on the data type.  In \cite{HenzingerKirschPayerSezginSokolova13}, Henzinger et al. formalized these relaxations of abstract data type specifications by increasing the set of legal sequences and defined several parameterized ways to do so.

These relaxations and other work which followed \cite{TalmageWelch14, ShavitTaubenfeld16, TalmageWelch18} concentrated on relaxing sequential data type specifications and showed that they can be implemented more efficiently in a message passing system than an unrelaxed queue.  This approach cannot consider concurrency, which is simply not defined in the sequential space, so Casta\~{n}eda et al. \cite{CastanedaRajsbaumRaynal20} defined multiplicity queues, which allow different behavior in the presence of concurrent operations than during sequential operation.  These data types cannot be linearizable, so in \cite{CastanedaRajsbaumRaynal20}, set-linearizability replaced linearizability as the consistency condition we seek to provide.

In shared memory models, multiplicity queues have a number of advantages over unrelaxed queues, and even other, sequential relaxations.  Casta\~{n}eda et al. showed that multiplicity queues can be implemented purely from $Read$/$Write$ registers, which is impossible for FIFO queues \cite{Herlihy91} and most previously-considered relaxed queues \cite{ShavitTaubenfeld16,TalmageWelch19}, as they have consensus number 2.  This suggests that they are an excellent, practical way to get queue-like behavior cheaply in shared memory systems.  Casta\~{n}eda and Pi\~{n}a \cite{CastanedaPina21} use multiplicity queues to provide the first work-stealing algorithms without strong synchronization requirements.  Johnen et al. \cite{JohnenKhattabiMilani22} considered the time complexity of shared memory implementations of queues, implementing multiplicity queues in $O(\log n)$ steps for each of $Enqueue$ and $Dequeue$, while the best previous algorithm took $O(\sqrt{n})$ steps \cite{KhanchandaniWattenhofer18}.

\section{Model and Definitions}

\subsection{System Model}
To have parameters we can use to prove lower bounds, we work in a partially synchronous model of computation.  Lower bounds in this model will also apply in asynchronous models, so a high lower bound here is still meaningful.  We work in the same system model as \cite{Talmage22} and its precedents, a partially-synchronous, message passing model without process failures used in the literature for various shared data type implementation algorithms and lower bounds.  There are $n$ processes, $\{p_0,\dots,p_{n-1}\}$, which participate in an algorithm implementing a shared memory object.  Each process provides an interface which allows a user to invoke operations on the simulated shared memory object and generates responses to those invocations.  Users can invoke operations at any time when their particular computing process does not have a \emph{pending} operation--an invocation which does not yet have a matching response.  Processes have local clocks running at the same rate as real time, but each potentially offset from real time, and can use these clocks to set timers.  

Processes are state machines, where each step is sending or receiving a message, setting or expiring a local timer, or local computation.  Operation invocations, message arrivals and timer expirations trigger steps of the state machine, which may perform local computation, set timers, send messages, and generate operation responses.  A \emph{run} of an algorithm is a set of sequences of state machine steps, one sequence for each process.  Each sequence in a run is a valid state machine history with a real time for each step, and is either infinite or ends in a state with no unexpired timers and no messages sent to that process but not received.  A run is \emph{admissible} if every message send step has a uniquely corresponding message receive step, and the \emph{delay} between send and receive is at least $d-u$ and at most $d$ real time, and the skew, or maximum difference between local clocks, is at most $\varepsilon := (1-1/n)u$ \cite{LundeliusLynch84}.   We assume $d$ and $u\leq d$ (and therefore $\varepsilon$) are known system parameters.

We are interested in data type implementations which satisfy certain conditions.  An implementation must provide \emph{liveness}, which means that every operation invocation must have a matching response.  We call this pair of invocation and corresponding response an \emph{operation instance}.  We are exploring the time cost of the implementation, as measured by the worst-case delay between an instance's invocation and response.  For an operation $OP$ specified in a data type, let $|OP|$ denote the maximum, over all admissible runs, of the difference in real time between the invocation and response of any instance of $OP$.  We are measuring communication cost, so we assume all local computation is instantaneous.  We also restrict ourselves to \emph{eventually quiescent} implementations, requiring that if there are a finite number of operation invocations in a run, there is a finite time after the last invocation by which the process will reach and stay in a \emph{quiescent} state with no messages in transit and no timers set.  A \emph{uniform} algorithm is one which is independent of the number of processes, so cannot change its logic based on different values of $n$.

A sequential data type specification gives a set of operations the user may invoke, with argument and return types, and the set of legal sequences of instances of those operations.  We are interested in data types whose behavior may depend on concurrency in a distributed system, so we consider \emph{set-sequential data type specifications}.  A set-sequential data type specification similarly defines the set of operation the user may invoke, but instead of a set of legal sequences of instances, specifies a set of legal sequences of \emph{sets of instances}.  Thus, not all instances in a run must be totally ordered relative to each other, but each set of instances must be totally ordered relative to others.

We are interested in \emph{set-linearizable} implementations of set-sequential data types, as defined in \cite{Neiger94} and \cite{CastanedaRajsbaumRaynal20}.  Set-linearizability requires that for every admissible run of the algorithm, there must be a total order of sets of operation instances which contains every instance in the run, is legal by the set-sequential data type specification, and respects the real-time order of non-overlapping instances.  That is, there must be a way to place all operation instances in the run in sets and order those sets into a legal sequence such that for every pair of instances where $op_1$ returns before $op_2$'s invocation, $op_1$ is in a set which precedes the set containing $op_2$.  The classic notion of linearizability is a special case of set-linearizability where all sets are required to have cardinality 1.

\subsection{Multiplicity Queues}

A \emph{queue} is a First-In, First-Out data type providing operations $Enqueue(arg)$ which returns nothing and $Dequeue()$ which returns a data value, where any sequence of instances of these operations is legal iff each $Dequeue$ instance returns the argument of the earliest preceding $Enqueue$ instance whose argument has not already been returned by a $Dequeue$, or the special value $\bot$ (which cannot be an $Enqueue$ instance's argument) if there is no such $Enqueue$ instance.  We consider a related data type called a \emph{multiplicity queue}, defined in \cite{CastanedaRajsbaumRaynal20}, which provides the same operations but is defined set-sequentially.

\begin{definition}
  A \emph{multiplicity queue} over value set $V$ is a data type providing two operations: $Enqueue(arg)$, which takes one data parameter $arg \in V$ and returns nothing, and $Dequeue()$, which takes no parameter and returns one data value in $V \cup\{\bot\}$, where $\bot$ is special value indicating that the structure is empty.  A sequence of sets of $Enqueue$ and $Dequeue$ instances is legal if (i) every $Enqueue$ instance is in a singleton set, (ii) all $Dequeue$ instances in the same set return the same value, and (iii) each $Dequeue$ instance $deq$ returns the argument of the earliest $Enqueue$ instance preceding $deq$ in the sequence, which has not been returned by another $Dequeue$ instance preceding $deq$.  If there is no such $Enqueue$ instance, $deq$ returns $\bot$.
\end{definition}

Note that the set of legal sequences of sets of operation instances must be prefix-closed.

This definition implies that in a set-linearizable implementation of a multiplicity queue, any two concurrent $Dequeue$ instance may, but do not necessarily, return the same value.  Such instances would be placed in the same set.  If two $Dequeue$ instances are not concurrent, then one must precede the other in the set linearization, so they must return different values.  Note that we will assume that all $Enqueue$ arguments are unique, which is easily achieved by a higher abstraction layer timestamping the user's arguments.  

\subsection{Shifting Proofs}

To prove our lower bounds, we will use indistinguishability arguments, where we argue that in a given time range in two runs, one or more processes has the same inputs (invocations, messages, timers) at the same local clock times.  Since each process is a (deterministic) state machine, a process which receives the same inputs at the same times must perform the same steps in the two runs.  We will sometimes argue the indistinguishability of two runs directly, but in some cases we will use \emph{shifting} \cite{LundeliusLynch84,Kosa99,WangTalmageLeeWelch18}.  Shifting is a technique which mechanically changes the real time at which each event at one or more processes occurs, while adjusting message delays and clock offsets to ensure that each event happens at the same local time at each process.  Thus, if one run is a shift of another, they are necessarily indistinguishable.  More formally, given run $R$ and vector $\vec{v}$ of length $n$, we define $Shift(R,\vec{v})$ as a new run in which each event $e$ at each $p_i, 0 \leq i < n$ which occurs at real time $t$ in $R$ occurs at real time $t+v[i]$.  To ensure that the runs are indistinguishable to the processes, local clock offsets $c_i$ are changed to $c_i' = c_i - v[i]$.  Finally, any message from $p_i$ to $p_j$ which had delay $x$ in $R$ has delay $x + v[j] - v[i]$, as the real times when it is sent and received change.

The primary challenge in using shifting arguments is that the new, shifted run must be admissible for us to require the algorithm to behave correctly.  Great care is required to define a run's message delays and clock offsets so that the skew and message delays are within the model's bounds after shifting.  Wang et al., in \cite{WangTalmageLeeWelch18}, extended the classic idea of shifting by showing that if a shift is too large, making the shifted run inadmissible, it is in some cases possible to chop off each process' sequence of steps before a message arrives after an inadmissible delay, then extend the run from that collection of chop points with different message delays which are admissible.  This extended run is not necessarily indistinguishable past the chop, but we can in some cases argue that the runs are indistinguishable long enough to form conclusions about the over-shifted run's behavior.  We use this technique to enable us to create a run, shift it too far to be inadmissible, modify it to a similar but admissible run, and argue that the two are indistinguishable.  Since this is not just a shift, this indistinguishability is not guaranteed, but the technique in \cite{WangTalmageLeeWelch18} allows us to argue indistinguishability to the point we need in each pair of runs.

\section{Lower Bound Proof Outline}

Our primary result is a lower bound on the worst-case time of any uniform set-linearizable implementation of a multiplicity queue.  This lower bounds shows that any possible implementation of a multiplicity queue, even in a relatively friendly model such as our partially synchronous, failure-free one, will be expensive.  For comparison, a linearizable implementation of an unrelaxed FIFO queue is possible with worst-case $Dequeue$ cost $d+\varepsilon = d+(1-1/n)u$.  Our bound is over half of that, indicating a limit on the performance gains of a multiplicity queue versus a traditional one.  Since our lower bound shows the impossibility of a more efficient multiplicity queue implementation in a relatively well-behaved partially synchronous model of computation, it follows that it is similarly impossible in more realistic, and less well-behaved, models, such as those which all asynchrony or failures.

We prove our bound by building up two sets of runs.  In both sets, each process invokes a single $Dequeue$ instance.  In the first set we show that each of these $Dequeue$ instances, despite being concurrent with at least one other $Dequeue$ instance, returns a unique value.  In the second set, we show that there are fewer distinct return values than $Dequeue$ instances, so there must be some pair of $Dequeue$ instances returning the same value.  We then show that, for sufficiently large $n$, these two sets of runs eventually converge, in the sense that processes cannot distinguish which set they are in until after they must choose return values for their $Dequeue$ instances. This means they must have the same behavior in both which contradicts our assumption on the worst-case cost of $Dequeue$.  We need large $n$ to ensure that the information about all of the $Dequeue$ instances cannot reach the last process in time for it to distinguish which run it is in.

Both sets of runs we use are based on and building towards one simple run, which sequentially enqueues values $1..n$, then has each process dequeue one value, with invocation times staggered so that the $Dequeue$ instances at different processes overlap slightly.  The idea is that processes each invoke $Dequeue$ slightly before the previous process' $Dequeue$ must have returned.  One complication is that if $|Dequeue| < u$, the math for invocation timing would have later processes invoking $Dequeue$ earlier, which we do not want, so in that case we have all processes invoke $Dequeue$ at the same time.  We use a variable $s$ to handle the different timings for these two cases.  

Every run we will use will start with process $p_0$ sequentially executing the sequence $Enqueue(1) \cdot Enqueue(2) \cdots Enqueue(n)$.  Then nothing happens until such time as the algorithm becomes quiescent, and fix a time $t_1$ after that point.  Thus, any set linearization of any of our runs will start with $n$ singleton sets, enqueueing the values $1..n$ in order.  All further operation instances will set-linearize after those $Enqueue$ instances.  In general in our runs, messages from lower-indexed processes to higher-indexed processes take $d-u$ time, while those from higher-indexed processes to lower-indexed processes take $d$ time.  The primary exception is that after a certain point, messages from $p_{n-1}$ to $p_n$ will also take $d$ time.  As we develop our proof, we will also have other delays, but all defined from this pattern. This prevents $p_n$ from collecting complete information on the previous portion of the run, which we will show is enough uncertainty to cause incorrect behavior.  

Let $Q := \min\left\{\frac{3d+2u}{5}, \frac{d}{2}+u\right\}$ throughout the paper.  We will assume $|Dequeue| < Q \leq d$.

\section{Distinct Return Values}

For our first set of runs, we construct a sequence of runs as outlined above and show that each $Dequeue$ instance may return a distinct value, despite the fact that each is concurrent with at least one other instance.  While this is the easier part of the proof, it is interesting as it shows that, under uncertainty in message delay, processes cannot tell whether their $Dequeue$ instances are or are not concurrent, so the relaxation gives no advantage, as processes must spend time to choose distinct return values.

We denote this set of runs by $D_k, 1\leq k \leq n$, where the first $k$ processes invoke $Dequeue$ instances slightly overlapped as discussed, and higher-indexed processes invoke $Dequeue$ slightly later.  We will inductively show that the $Dequeue$ at $p_k$ must return a different value from those at $p_0,\dots,p_{k-1}$, then shift the run to obtain $D_{k+1}$, which is indistinguishable.  When the inductive chain of shifts is complete, we will see that all $n$ $Dequeue$ instances in $D_n$ must return different values.

\begin{construction}
  Define run $D_k$ ($D$ for \underline{D}istinct) as follows, for each $1\leq k < n$:
  \begin{itemize}
  \item $p_0$ invokes $Enqueue(1) \cdots Enqueue(n)$ in order.  Let $t_1$ be an arbitrary time after $Enqueue(n)$ returns at which the system is quiescent.
  \item $\forall 0 \leq i < k$, process $p_i$ invokes $Dequeue$ at time $t_1 + i*s$, where $s = \max\{0,Q - u\}$.
  \item $\forall k\leq j < n$, process $p_j$ invokes $Dequeue$ at time $t_1 + (j-1)s + (s+u)$.
  \item Process $p_0$ has local clock offset $c_0 = 0$.
  \item $\forall 0 < i < k$, process $p_i$ has local clock offset $c_i = \left(\frac{i}{n}\right)u$.
  \item $\forall k \leq j < n$, process $p_j$ has local clock offset $c_j = \left(\frac{j-n}{n}\right)u$.
  \item $\forall 0 \leq i < k \leq j < n$, messages from $p_i$ to $p_j$ have delay $d$, from $p_j$ to $p_i$ have delay $d-u$.
  \item $\forall 0 \leq a < b < k$, messages from $p_a$ to $p_b$ have delay $d-u$, from $p_b$ to $p_a$ have delay $d$.
  \item $\forall k \leq c < d < n$, messages from $p_c$ to $p_d$ have delay $d-u$, from $p_d$ to $p_c$ have delay $d$.
  \end{itemize}
  Define run $D_n$ identically for all processes $p_i$ with $i<n$.  Since $p_n$ does not exist, it does not invoke $Dequeue$, send or receive messages, or have a local clock offset.
\end{construction}

Since all local clock offsets for processes $p_i$ with $0 < i < k$ are positive and increase with $i$ and all offsets for processes $p_j$ with $k \leq j < n$ are negative and increase with $j$, the maximum skew between processes is $|c_{k-1} - c_{k}| = \left|\frac{k-1}{n} - \frac{k-n}{n}u\right| = \frac{n-1}{n}u = \varepsilon$, except when $k = n$, when no such $p_i$ exists and the maximum skew is $\left|0 - \frac{n-1}{n}u\right| = \varepsilon$.  With this fact and since all message delays are in the range $[d-u,d]$, we see that each $D_k$ is an admissible run.

\begin{lemma}
  For all $2 \leq k < n$, $D_k = Shift(D_{k-1},\overrightarrow{s_{k-1}})$, where $\overrightarrow{s_{k-1}}$ is the 0 vector, except that the value at index $k-1$ is $-u$: $\overrightarrow{s_{k-1}} = \langle 0,\dots,0,-u,0,\dots,0\rangle$.
\end{lemma}

\begin{proof}
  Let $k$ be an arbitrary value with $2 \leq k < n$.  Consider what happens when we shift $D_{k-1}$ by $\overrightarrow{s_{k-1}}$.  All events at $p_{k-1}$ occur $u$ earlier in real time, so $p_{k-1}$ invokes $Dequeue$ at time $t_1 + ((k-1)-1)s + (s+u) - u = t_1 + (k-2)s + s = t_1 + (k-1)s$, which matches the definition of $D_k$.  Let $0 \leq i < k-1 < j < n$.  Message delays in $D_{k-1}$ from $p_{k-1}$ to $p_i$ were $d-u$, and from $p_i$ to $p_{k-1}$ were $d$.  In the other direction, messages delays from $p_{k-1}$ to $p_j$ were $d-u$ and from $p_j$ to $p_{k-1}$ were $d$.  When we shift the send and receive events at $p_{k-1}$ earlier, messages from $p_{k-1}$ have a longer delay by $u$ and messages to $p_{k-1}$ have a shorter delay $u$.  We see that this leaves all delays from $p_{k-1}$ to another process at $d$ and all delays to $p_{k-1}$ at $d-u$, which are admissible.  Since we only shifted one process, messages between other processes are unchanged.

  Finally, we consider clock offsets.  $c_{k-1}$ is $\left(\frac{(k-1)-n}{n}\right)u$ in $D_{k-1}$, and must increase by $u$ to hide the difference in real time when we shift.  Thus, in $Shift(D_{k-1}, \overrightarrow{s_{k-1}})$, $c_{i-1} = \left(1 + \frac{(k-1)-n}{n}\right)u = \left(\frac{(k-1)}{n}\right)u$, matching the specification for $D_k$.
\end{proof}

\begin{lemma}\label{lem:distinctReturns}
  In run $D_k$, $1 \leq k \leq n$, every $Dequeue$ instance returns a distinct value.  Specifically, for each $0 \leq i < n$, the $Dequeue$ instance at $p_i$ returns $i+1$.
\end{lemma}

\begin{proof}
  We proceed by induction on $k$, from $1$ to $n$.
  
  Base Case: Consider $D_1$.  Here, $p_0$ invokes $Dequeue$ at time $t_1$, which must return by time $t_1 + |Dequeue|$.  $p_1$ invokes $Dequeue$ at time $t_1 + (1-1)s + (s+u) > t_1 + |Dequeue|$, which is after $p_1$'s $Dequeue$ instance returns.  Every process $p_i$ with $i \geq 1$ invokes $Dequeue$ no earlier than $p_1$, so no other $Dequeue$ instance is concurrent with $p_0$'s, and thus that one must set-linearize before any other.  This means that $p_0$ returns $1$ to its $Dequeue$ instance and all other processes return values in the set $\{2,\dots,n\}$ to their $Dequeue$ instances.

  Inductive Hypothesis: Assume that for some arbitrary $0 \leq k < n-1$, each process $p_i$, $0\leq i < k$ returns $i+1$ to its $Dequeue$ instance.  

  Inductive Step: We will show that process $p_k$ returns $k+1$ to its $Dequeue$ instance.  First, note that in $D_k$, $p_k$ invokes $Dequeue$ at time $t_1 + (k-1)s + (s+u)$, while every $p_i, 0\leq i < k$ has its $Dequeue$ instance return no later than $t_1 + (i-1)s + |Dequeue| \leq t_1 + ((k-1)-1)s + |Dequeue| < t_1 + (k-2)s + (s+u)$.  Since $s \geq 0$, this is before $p_k$ invokes $Dequeue$, so $p_k$'s $Dequeue$ instance must set-linearize strictly after all of those at any lower-indexed $p_i$.  By the inductive hypothesis, each of those $k$ processes returns $i+1$, so $p_k$ must return a value larger than $k$.

  Now, consider $D_{k+1}$.  Since $D_{k+1}$ is a shifted version of $D_k$, no process can distinguish the two runs, so all behave the same way in both.  Specifically, $p_k$ will return the same value to its $Dequeue$ instance.  But in $D_{k+1}$, by an identical argument to that in the previous paragraph, each $p_j, k< j < n$ invokes $Dequeue$ after the $Dequeue$ instance at $p_k$ returns, so they must all set-linearize strictly after $p_k$'s $Dequeue$ instance, and those of each $p_i, 0 \leq i \leq k$.  Since there are only $k+1$ $Dequeue$ instances set-linearized with or before that at $p_k$, these must return values from the set $\{1,\dots,k+1\}$.  But we know that those at processes with indices in $\{0,\dots,k-1\}$ all return values from $\{1,\dots,k\}$, and the $Dequeue$ instance at $p_k$ returns a value distinct from any of these, so it must return $k+1$, and we have the claim.
\end{proof}

\section{Repeated Return Values}

For the second set of runs, we will show that $n$ processes, each invoking one $Dequeue$ instance in our same partially-overlapping pattern, will not return all different values to those $Dequeue$ instances.  To do this, we first show that if only three processes invoke $Dequeue$, then they will only return two distinct values.  We then inductively construct more and more complex runs, with one more process joining the pattern and invoking $Dequeue$ in each successive pair of runs.  When the induction reaches $n$, we will show that we have a run indistinguishable from the $D_n$ we constructed in the previous section.  Since each of the $Dequeue$ instances in that run returns a distinct value, and those in the run we construct here do not all return distinct values, we have a contradiction, proving that the assumed algorithm cannot exist.

First, we define the family of runs $S_k$ in each of which only $k \leq n$ processes invoke $Dequeue$.  We will inductively show that each of these has some pair of $Dequeue$ instances which return the same the same value, eventually showing that not all $Dequeue$ instances in $S_n$ return distinct values.

\begin{construction}
  Define run $S_k$ ($S$ for \underline{S}ame, since there are $Dequeue$ instances with the same return value) as follows:
  \begin{itemize}
  \item $p_0$ invokes $Enqueue(1) \cdots Enqueue(n)$ in order.  Let $t_1$ be the same arbitrary time after $Enqueue(n)$ returns at which the system is quiescent as in the definition of $D_k$.
  \item $\forall 0 \leq i < k$, process $p_i$ invokes $Dequeue$ at time $t_1 + i*s$, where $s = \max\{0,Q - u\}$.
  \item Process $p_0$ has local clock offset $c_0 = 0$, and $\forall 0 < i < n$, process $p_i$ has $c_i = \left(\frac{i}{n}\right)u$.
  \item $\forall 0 \leq i < j < n$, messages from $p_j$ to $p_i$ have delay $d$ and from $p_i$ to $p_j$ have delay $d-u$, except for those from $p_{k-2}$ to $p_{k-1}$ sent after $t_{k-2}^* = t_1 + (k-2)(d-u)$, which have delay $d$.
  \end{itemize}
\end{construction}

To show the chain of indistinguishabilities in our induction, we will need another set of runs, which are intermediate steps.

\begin{construction}
  For $1 \leq k < n$, define run $S_k'$ from run $S_{k-1}$ by additionally having $p_{k-1}$ invoke $Dequeue$ at time $t_1 + (k-1)s$.  Adjust the delay of all messages from $p_{k-2}$ to $p_{k-1}$ sent at or after $t_{k-2}^* = t_1 + (k-2)(d-u)$ to $d$.
\end{construction}

In $S_k'$, we have added the next $Dequeue$ instance, but have two processes' messages ($p_{k-3}$'s and $p_{k-2}$'s) to the next, larger-indexed, process delayed.  We can show that processes $p_0$ through $p_{k-1}$ cannot distinguish $S_{k-1}$ from $S_k'$ before generating return values for their $Dequeue$ instances, so they must return the same values, which gives us information about what $p_k$ must return to its $Dequeue$ instance.  We then show that $S_k'$ and $S_k$ are indistinguishable to $p_k$ until after it has generated a return value for its $Dequeue$ instance, telling us what values it could return.   

\begin{lemma}\label{lem:sameReturns}
In $S_n'$ and $S_n$, all $Dequeue$ instances return values from the set $\{1,\dots,n-1\}$, for sufficiently large $n$.
\end{lemma}

\begin{proof}
We proceed by mathematical induction on $k$, from $3$ to $n$.  

\subsection*{Base Case}\label{sec:sameValuesBaseCase}
\begin{claim}
In run $S_3$, all $Dequeue$ instances return values from the set $\{1,2\}$, for sufficiently large $n$.
\end{claim}

We start with only the first three of our $n$ processes invoking $Dequeue$, which is run $S_3$.  Due to higher-indexed processes invoking $Dequeue$ later than lower-indexed processes, and the way we will set message delays, the first $Dequeue$ instance will behave as if it were running alone, returning $1$.  We will then shift run $S_3$, using a technique like that in \cite{WangTalmageLeeWelch18} that allows us to over-shift and break some message delays, then re-insert those messages with new, admissible delays.  We can then show that the resulting patched run is still indistinguishable from the starting run for long enough.  In this run, we will argue that the second process does not learn about the first process' $Dequeue$ instance until after its own returns, and thus cannot distinguish this run from one in which it is running alone, so it must also return 1.  Given these two return values, set-linearizability implies that the third process' $Dequeue$ instance must return 2.  We will then show that the third process cannot distinguish between the original and shifted runs before choosing its return value, so will return $2$ in $S_3$.

\begin{proof}
  First, observe that $p_0$ cannot learn about the $Dequeue$ instances at $p_1$ and $p_2$ until after its own $Dequeue$ instance has returned.  Since all messages from a higher-indexed process to a lower-indexed process have delay $d$, the earliest $p_0$ will learn about the other $Dequeue$ instances is at time $t_1 + s + d$, since $t_1 + s$ is when $p_1$ invokes $Dequeue$, and any message indicating that this has happened will take $d$ time to reach $p_0$.  Since $p_2$ invokes its $Dequeue$ instance at time $t_1 + 2s \geq t_1 + s$, the same logic will imply that $p_0$ will also not learn about that instance until after its own $Dequeue$ instance has returned.  $p_0$'s $Dequeue$ instance returns no later than time $t_1 + |Dequeue|$, by definition, which is strictly less than $t_1 + d$.  Together, we see that $p_0$ learns about a remote $Dequeue$ invocation no sooner than $t_1 + s + d  \geq t_1 + d > t_1 + |Dequeue|$, so through the return of its $Dequeue$ instance, $p_0$ cannot distinguish $S_3$ from a run in which that is the only $Dequeue$ instance.  Thus, it returns the same value, which by set-linearization is necessarily $1$.  Similarly, $p_1$ must return a value in $\{1,2\}$, since it cannot learn about the $Dequeue$ instance at $p_2$ until time $t_1 + 2s + d$, which is larger than when its own $Dequeue$ instance returns by $t_1 + s + |Dequeue|$.

  Next, we want to show that $p_2$ will also return a value from $\{1,2\}$ to its $Dequeue$ instance.  We cannot directly argue this, since if $p_2$ learns about both the $Dequeue$ instances at $p_0$ and $p_1$ before it generates a return value for its own, it may decide to return a different value than either.  Instead, we will shift events at $p_1$ earlier, then argue that in this run, the information about $p_0$'s $Dequeue$ instance does not arrive at $p_1$ until after it has generated its $Dequeue$ return value, forcing $p_1$ to return 1 to its $Dequeue$ instance.  Now, while $p_1$ may be able to distinguish this new run from $S_3$, we will argue that $p_2$ will not be able to distinguish them until after it generates its $Dequeue$ return, so must return the same value in both.  In the shifted run, $p_0$ and $p_1$ will both return 1, which means that $p_2$ must return either 1 or 2 to satisfy set-linearizability.

  We will shift $S_3$ by the vector $\langle 0, -X, 0, \dots, 0\rangle$, where $X$ is a value we will determine shortly.  Next, we will alter message delays, both to delay $p_1$ from learning about $p_0$'s $Dequeue$ instance and to make the run admissible again.  Call this new shifted and modified run $S_3^X$.

  \begin{table}
    \centering
    \caption{Table showing message delays to and from $p_1$ in runs for base case of overlapping $Dequeue$ return values proof.  Only delays to/from $p_1$ appear, since all others are unchanged across the three runs.}\label{table:baseCaseDelays}
    \begin{tabular}{|l|c|c|c|}\hline
      Message Path                  & $S_3$ & $Shift(S_3)$                         & Adjusted: $S_3^X$ \\\hline
      $p_0 \to p_1$                 & $d-u$ & \cellcolor{red!30!white}$d-u-X$      & \cellcolor{blue!30!white}$d$ \\\hline
      $p_1 \to p_0$                 & $d$   & \cellcolor{red!30!white}$d+X$        & \cellcolor{blue!30!white}$d$ \\\hline
      $p_1 \to p_2$ (initially)     & $d-u$ & $d-u+X$                              & $d-u+X$ \\\hline
      $p_1 \to p_2$ (after $t_1^*$) & $d$   & \cellcolor{red!30!white}$d+X$        & \cellcolor{blue!30!white}$d$ \\\hline
      $p_1 \to p_{\geq 3}$          & $d-u$ & $d-u+X$                              & $d-u+X$ \\\hline
      $p_2 \to p_1$                 & $d$   & $d-X$                                & $d-X$ \\\hline
      $p_{\geq 3} \to p_1$          & $d$   & $d-X$                                & $d-X$ \\\hline
    \end{tabular}
  \end{table}

  Our first step is to find what shift amounts $X$ for $p_1$ will make $S_3^X$ admissible, then argue the behavior of each process.  First, note that this shift will increase the local clock offset of $p_1$ by $X$.  In $S_3$, $c_1 = \frac{1}{n}u$, the smallest clock offset is $c_0 = 0$ and the largest is $c_{n-1} = \frac{n-1}{n}u$.  To keep the run admissible, we must have $X \leq \left(\frac{n-1}{n}u - \frac{1}{n}u\right) = \frac{n-2}{n}u$, since we are not changing the smallest clock offset, so must keep $c_1$ within $\varepsilon$ of that offset.

  Next, we see that for a non-negative value of $X$, we will have some inadmissible message delays in $Shift(S_3, \langle 0,-X,0,\dots,0\rangle)$ (highlighted in red in the $Shift(S_3)$ column).  To correct these, we trim the run before any of the inadmissible messages would arrive, then extend the run with other, admissible message delays (highlighted in blue in the $S_3^X$ column), following the technique introduced in \cite{WangTalmageLeeWelch18}.  Unlike a shift, this may change the behavior of the run, so we will argue what each process does in run $S_3^X$.  The choice of these new delays is based on trying to delay processes from learning about remote actions, which is why we set all of the adjusted delays to the maximum, $d$.  Since $X \leq \frac{n-2}{n}u < u$, then the delays not highlighted in Table~\ref{table:baseCaseDelays} are in the range $[d-u,d]$, and we conclude that if $0 \leq X \leq \frac{n-2}{n}u$, then $S_3^X$ is admissible.

  Now that we know what values of $X$ make $S_3^X$ an admissible run, we will find which of those values of $X$ will make all three $Dequeue$ instances return values from $\{1,2\}$ in $S_3^X$.

  $p_0$ will not learn about $p_2$'s $Dequeue$ instance until after its own returns, by the same argument as in $S_3$.  We want $p_0$ to also not learn about $p_1$'s $Dequeue$ instance until after its own returns.  $p_1$ invokes $Dequeue$ at $t_1 + s - X$ in $S_3^X$, since we shifted events at $p_1$ earlier by $X$.  A message sent at this time will arrive at $p_0$ at time $t_1 + s - X + d$, and we want to argue that this will be after $t_1 + |Dequeue|$, and thus after $p_0$'s $Dequeue$ instance returns.  This happens if and only if $d + s - X > |Dequeue|$, or $X < d + s - |Dequeue|$.  Since $s \geq 0$, it is sufficient to require that $X < d-|Dequeue|$ to ensure that $p_0$'s $Dequeue$ instance returns $1$.

  To force $p_1$'s $Dequeue$ instance to return 1, we want information about $p_0$'s invocation of $Dequeue$ to arrive after $p_1$ generates its $Dequeue$ return value.  Thus, we want to have time $t_1 + d$ (since messages from $p_0$ to $p_1$ have delay $d$ in $S_3^X$) later than when $p_1$ generates a return value.  $p_1$ invokes $Dequeue$ at time $t_1 + s - X$ and the $Dequeue$ instance returns at most $|Dequeue|$ time after invocation, so we want to have $t_1 + d > t_1 + s - X + |Dequeue|$.  Solving for $X$, we find that this is true iff $X > |Dequeue| + s - d$.  Here, we split into cases depending on the value of $s$:
  \begin{itemize}
  \item $s = 0$: Then we want $X > |Dequeue|-d$, but we assumed that $|Dequeue| < d$, so any non-negative value of $X$ is sufficient.
  \item $s = Q-u$: Then we want $X > |Dequeue| + (Q-u) - d < |Dequeue| + Q - (d+u)$.  
  \end{itemize}

  Similarly to previous arguments, since $p_2$ invokes $Dequeue$ at least $X$ after $p_1$ does ($p_2$ invokes $Dequeue$ $s$ after $p_1$ in $S_3$, which means $s+X$ after in $S_3^X$), and message delays from $p_2$ to $p_1$ are $d-X$, $p_1$ cannot learn about $p_2$'s $Dequeue$ invocation until at least $X + (d-X) = d > |Dequeue|$ after $p_1$ invokes $Dequeue$.  This is after $p_1$ generates its $Dequeue$ return value.  Combining this with the previous conclusion that $p_1$ is unaware of $p_0$'s $Dequeue$ invocation until after it chooses a return value, we conclude that $p_1$ will return the same value as in a run where neither $p_0$ nor $p_2$ invoked $Dequeue$.  The only legal set-linearization of such a run requires that $p_1$ return 1.

  We can now reason about $p_2$'s behavior.  Since both $p_0$ and $p_1$ must return 1 to their $Dequeue$ instances in $S_3^X$, we conclude that $p_2$ must return either $1$ or $2$, as there is no legal set-linearization of any other return value.  We will thus argue that $p_2$ cannot distinguish $S_3^X$ from $S_3$ until after it generates its $Dequeue$ return value, concluding that $p_2$ will return either 1 or 2 to its $Dequeue$ instance in $S_3$ as well.

  Consider when each process in $S_3$ can first distinguish that it is not in $S_3^X$.  These differences correspond to the adjusted message delays highlighted in the final column of Table~\ref{table:baseCaseDelays}.  $p_0$ can distinguish the runs when it does not receive a message $p_1$ may have sent at its $Dequeue$ invocation as soon as it would have received it in $S_3^X$, since in $S_3^X$ we reduced the delay on messages from $p_1$ to $p_0$.  This detection would occur at time $t_1 + s + (d-X)$, when that message does not arrive.  $p_1$ can first distinguish the runs at time $t_1 + (d-u)$, when it can receive a message $p_0$ sent at its $Dequeue$ invocation but which arrives later in $S_3^X$, where we increased the delay on messages from $p_0$ to $p_1$.  Note $t_1 + s + (d-X) + (d-u) > t_1 + (d-u)$ and $t_1 + s + (d-X) \leq t_1 + (d-u) + d$, so neither process can send a message after it detects the difference which will arrive before the recipient detects the difference itself.

  Finally, $p_2$ can distinguish the runs either by receiving a message $p_0$ or $p_1$ sends after distinguishing the runs or directly from adjusted message delays.  We argue that each of these must occur after the $Dequeue$ instance at $p_2$ returns, so $p_2$ cannot distinguish $S_3$ from $S_3^X$ until after that $Dequeue$ instance's return value is set, so the value must be the same in both runs.

  Consider when $p_2$ can receive a forwarded detection of a difference in the runs:
  \begin{itemize}
  \item $p_0$ can send this information no sooner than $t_1 + s + (d-X)$, and the message would take $d$ time to arrive, meaning that the earliest $p_2$ could distinguish the runs based on this information is $t_1 + s + 2d - X$.  We want to show that this is greater than $t_1 + 2s + |Dequeue|$, and thus after $p_2$'s $Dequeue$ instance returns.  This is true iff $2d - X - u > s + |Dequeue|$.  Consider cases for the value of $s$:
    \begin{itemize}
    \item $s=0$: We want to show that $2d - X - u > |Dequeue|$.  This is true if and only if $X < (d - |Dequeue|) + (d - u)$, but we know that $d \geq u$ so this holds if $X < d - |Dequeue|$.
    \item $s = Q - u$: We want to show that $2d - X - u > |Dequeue| + Q - u$.  This is true if and only if $X < (d-Q) + (d-|Dequeue|)$.  Since $Q \leq d$ and we are already assuming $X < d-|Dequeue|$, this inequality holds.
    \end{itemize}
  \item $p_1$ can send a message informing $p_2$ that it is in $S_3$, not $S_3^X$, no sooner than $t_1 + (d-u)$.  Since $t_1 + (d-u) = t_2^*$, this message will take $d$ time to arrive at $p_2$.  We want to show that this is after $p_2$'s $Dequeue$ instance returns, which happens at $t_1 + 2s + |Dequeue|$.  Thus, we want $t_1 + (d-u) + d > t_1 + 2s + |Dequeue|$, or $2d-u > 2s + |Dequeue|$.  Solving for $|Dequeue|$, this is equivalent to $|Dequeue| < 2d-2s-u$.  Consider the possible values of $s$:
    \begin{itemize}
    \item $s = 0$: We want to show that $|Dequeue| < 2d-u$.  But we know that $d\geq u$, so $d-u \geq 0$ and $|Dequeue| < d$, so this inequality holds.
    \item $s = Q - u$: We want to show that $|Dequeue| < 2d - 2(Q-u) - u = 2d - 2Q + u$.  But $|Dequeue| < Q$, so it is sufficient to show that $Q \leq 2d-2Q+u$.  This holds iff $Q \leq \frac{2d+u}{3}$.  But we assumed $Q \leq \frac{3d+2u}{5} \leq \frac{2d+u}{3}$, so we have the desired relationship.
    \end{itemize}
    Thus, $p_2$ cannot learn from $p_1$ that it is in $S_3^X$ before it generates a return value for its $Dequeue$ instance.
  \end{itemize}

  Now, we show that $p_2$ cannot directly differentiate $S_3$ from $S_3^X$ based on the altered message delays in $S_3^X$ before its $Dequeue$ instance returns.  At the earliest, this can happen at $t_2^* + d - X$, when $p_2$ does not receive a message in $S_3$ that it may have in $S_3^X$, since in $S_3^X$ we decreased the delay of messages $p_1$ sends to $p_2$ at or after time $t_2^*-X$.  We again want to show that this is after $p_2$'s $Dequeue$ instance returns which happens no later than $t_1 + 2s + |Dequeue|$.  That is, we want $t_2^* + d - X = t_1 + (d-u) + d - X > t_1 + 2s + |Dequeue|$.  Equivalently, we want $2d-u-X > 2s + |Dequeue|$.  Consider cases for $s$:
  \begin{itemize}
  \item $s = 0$: In this case, we want $2d-u-X > |Dequeue|$, which is true iff $X \leq (d-|Dequeue|) + (d-u)$.  We already have the constraint that $X < d-|Dequeue|$ and $u \leq d$, so this is true.
  \item $s = Q - u$: Here, we want $2d-u-X > 2(Q-u) + |Dequeue|$, which is true if $X < 2d + u - 2Q - |Dequeue|$.  This is a new constraint on $X$ which we must meet to have the desired behavior.
  \end{itemize}

  Thus, in all cases (if $X$ meets all our constraints simultaneously), $p_2$ cannot distinguish $S_3$ from $S_3^X$ until after its $Dequeue$ instance returns.  This means that it returns the same value in both runs, and we proved that it must return a value from $\{1,2\}$ in $S_3^X$, so it does in $S_3$, as well.

  Our last step is to verify that our constraints on $X$ are compatible--that there is a value of $X$ which will make $S_3^k$ admissible and give the behavior we want.  Our constraints are
  \begin{itemize}
  \item $X \geq 0$ and $X > |Dequeue| + Q - (d+u)$
  \item $X < d-|Dequeue|$, $X < 2d + u - 2Q - |Dequeue|$, and $X \leq \frac{n-2}{n}u$
  \end{itemize}

  These three upper bounds and two lower bounds lead to six cases to check to show that there exists a value of $X$ which satisfies all of our constraints.
  \begin{itemize}
  \item Show that $d-|Dequeue| > 0$: By assumption, $|Dequeue| < d$, so $d-|Dequeue| > 0$.
  \item Show that $d-|Dequeue| > |Dequeue| + Q - (d+u)$: This is true iff $2d + u > 2|Dequeue| + Q$.  Since $|Dequeue| < Q$, it is sufficient to show that $2d+u \geq 3Q$, or $Q \leq \frac{2d+u}{3}$, but we assumed that $Q \leq \frac{3d+2u}{5} \leq \frac{2d+u}{3}$, so this relationship holds.
  \item Show that $2d+u-2Q - |Dequeue| > 0$: This is true iff $2d+u > 2Q + |Dequeue|$.  Again, it is sufficient to show that $2d+u \geq 3Q$, which is the same as the previous case.
  \item Show that $2d+u-2Q - |Dequeue| > |Dequeue| + Q - (d+u)$: This is true iff $3d + 2u > 3Q + 2|Dequeue|$, but it is sufficient to show that $3d+2u \geq 5Q$, and we assumed that $Q \leq \frac{3d+2u}{5}$, so this relationship holds.
  \item Show that $\frac{n-2}{n}u \geq 0$: $u \geq 0$, so any fraction of it will also be non-negative.
  \item Show that $\frac{n-2}{n}u > |Dequeue| + Q - (d+u)$: Solving for $|Dequeue|$ and $Q$, this is true iff $Q + |Dequeue| < d + \frac{n-2}{n}u + u$.  Since $|Dequeue| < \frac{d}{2} + u$, there is some $N_0$ s.t. for all $n\geq N_0$, $|Dequeue| < \frac{d}{2} + \frac{n-2}{n}u$.  Further $|Q| \leq \frac{d}{2} + u$, so together for $n\geq N_0$, the inequality holds for sufficiently large $n$.
  \end{itemize}
  Thus, since every upper bound is larger than every lower bound, for sufficiently large $n$ ($n\geq N_0$) there exists at least one $X$ such that $S_3^X$ is admissible and $p_0$, $p_1$, and $p_2$ all return values from $\{1,2\}$ to their $Dequeue$ instances, and we have the claim. 
\end{proof}

We will next proceed with the inductive case, showing that for each value of $k$, two processes in $S_k$ return the same value.  First, it is worth noting that, while it appears that we already have a contradiction by comparing $D_3$ and $S_3$, since these are very similar runs with different return values.  However, the runs differ in that in $S_3$, messages from $p_1$ to $p_2$ send at or after $t_1^*$ have delay $d$, while similar messages in $D_3$ would have delay $d-u$.  This means that we cannot argue that the runs are indistinguishable.  For that argument, in Section~\ref{sec:contradiction} below, we need a sufficiently large $n$ that in $D_n$, messages from $p_{n-2}$ to $p_{n-1}$ sent at or after $t_{n-2}^*$ do not arrive until after $p_{n-1}$'s $Dequeue$ instance returns, so that changing their delay cannot change that instance's return value.  Thus, for sufficiently large $n$, $S_n$ will be indistinguishable from $D_n$ until after all $Dequeue$ instances return and we will have our contradiction, but this does not necessarily hold for $n=3$.

\subsection*{Inductive Case}

Inductive Hypothesis:  Assume that for some arbitrary $4 \leq k \leq n$, all $Dequeue$ instances in $S_{k-1}$ return values from the set $\{1,\dots,k-2\}$.

Inductive Step: We will show that in $S_k$ and $S_k'$, all $Dequeue$ instances return values from the set $\{1,\dots,k-1\}$.  First, we will use $S_{k-1}$ to argue the behavior of $S_k'$, then use that behavior to prove the behavior of $S_k$.

To show that in $S_k'$, all processes return values from the set $\{1,\dots,k-1\}$ to their $Dequeue$ instances, we argue that processes $p_0, \dots, p_{k-1}$ cannot distinguish $S_{k-1}$ from $S_k'$ until after they have all generated their $Dequeue$ return values.  Thus, they will return the same values as in $S_{k-1}$, which are all in $\{1,\dots k-2\}$ by the inductive hypothesis.  We can then conclude that $p_{k-1}$, which invokes a $Dequeue$ instance in $S_k'$ but not in $S_{k-1}$ must return a value in the set $\{1,\dots,k-1\}$ to satisfy set-linearizability.

Recall that $S_k'$ differs from $S_{k-1}$ in two ways: First, $p_{k-1}$ invokes $Dequeue$ at time $t_1 + (k-1)s$.  Second, messages from $p_{k-2}$ to $p_{k-1}$ sent at or after $t_{k-2}^* = t_1 + (k-2)(d-u)$ have delay $d$ instead of $d-u$.  Thus, the first point at which any process can discern that it is in $S_k'$ instead of $S_{k-1}$ is $p_{k-1}$ at whichever of these events happens first.  For any other process, the first point where it can distinguish the runs is when it can receive a message $p_{k-1}$ sends after it discerns the difference.  We will argue that such a message arrives at any of $p_0,\dots,p_{k-2}$ after it has chosen a return value for its $Dequeue$ instance.  Note that we need only prove that such a message arrives at $p_{k-2}$ more than $|Dequeue|$ after it invokes $Dequeue$, since each process with a lower index invokes $Dequeue$ at the same time or sooner, and the message delay from $p_{k-1}$ to any lower-index process is the same.  We proceed by cases on which distinguishing event at $p_{k-1}$ occurs first.
\begin{itemize}
\item $p_{k-1}$ first distinguishes the runs when it invokes $Dequeue$:  The message delay from $p_{k-1}$ to $p_{k-2}$ is $d$, and any indirect path would take even longer, since any such path must have some message from a higher-indexed to lower-indexed process, which has delay $d$.  Thus, the earliest $p_{k-2}$ can distinguish the runs is $t_1 + (k-1)s + d$.  We want to show that this is later than the return time of $p_{k-2}$'s $Dequeue$ instance, which must return by $t_1 + (d-2)s + |Dequeue|$.  This inequality is true iff $t_1 + (k-1)s + d > t_1 + (k-2)s + |Dequeue|$, which reduces to $s + d > |Dequeue|$.
  
  Since $d > Q$ and $s \geq 0$, this inequality holds, which means that $p_{k-2}$ (and similarly $p_0,\dots,p_{k-3}$) cannot use the extra $Dequeue$ invocation at $p_{k-1}$ to distinguish $S_k'$ from $S_{k-1}$ until after their $Dequeue$ instances have returned.
\item $p_{k-1}$ first distinguishes the runs when it fails to receive a messages whose delay was increased:  The earliest possible sending time of such a message is $t_{k-2}^* = t_1 + (k-2)(d-u)$.  $p_{k-1}$ can detect that it has not arrived $d-u$ later (when it would have arrived in $S_{k-1}$), and then the earliest it can get information about the differentiation to a lower-indexed process is another $d$ after that.  We similarly want to show that this is after the $Dequeue$ instance at $p_{k-2}$ returns, which is true iff $t_1 + (k-2)(d-u) + (d-u) + d > t_1 + (k-2)s + |Dequeue|$, which reduces to $(k-1)(d-u) + d > (k-2)s + |Dequeue|$.

  Since $d > |Dequeue|$, $d\geq u$, and $s = \max\{0,Q-u\}$, we see that $d-u \geq s$, so the inequality holds.  Thus, in this case no process in $p_0,\dots, p_{k-2}$ can distinguish $S_k'$ from $S_{k-1}$ until after it has generated a return value for its $Dequeue$ instance.
\end{itemize}

Since in neither case can $p_0,\dots,p_{k-2}$ distinguish $S_k'$ from $S_{k-1}$ until after its $Dequeue$ instance returns, so all of those $Dequeue$ instances return the same values in both runs.  Specifically, by the inductive hypothesis they all return values from the set $\{1,\dots,k-2\}$.  The $Dequeue$ instance at $p_{k-1}$ must then return a value in the set $\{1,\dots,k-1\}$, as any larger value would violate set-linearizability, since there would be no $Dequeue$ instance returning $k-1$.

Now, having determined the behavior of $S_k'$, we use it to show that $S_k$ will behave similarly.  This is another indistinguishability argument, showing that $p_{k-1}$ cannot distinguish $S_k$ from $S_k'$, until after it has generated a return value for its $Dequeue$ instances.  Recall that the difference between $S_k$ and $S_k'$ is that in $R_k$ all messages from $p_{k-3}$ to $p_{k-2}$ have delay $d-u$, while in $S_k'$, those sent at or after $t_{k-3}^*$ have delay $d$.

Before we start the indistinguishability argument, note that if $p_k$ did not invoke $Dequeue$ in $S_k$, the remaining $k-1$ $Dequeue$ instances must return values from the set $\{1,\dots,k-1\}$, since there would only be $k-1$ instances, so there would be no way to set-linearize an instance that returned a larger value.  These processes must behave the same way in $S_k$ as in this run, since the first point where any could detect a difference would be $d$ after $p_k$'s invocation, which is after all other $Dequeue$ instances have returned, similar to prior arguments.  Thus, we need only concern ourselves with showing that $p_{k-1}$ cannot distinguish $S_k$ from $S_k'$ before its $Dequeue$ instance returns, so that it will return a value in $\{1,\dots,k-1\}$, as we proved it does in $S_k'$.

The only process which can directly detect a difference between $S_k$ and $S_k'$ is $p_{k-2}$, when it receives a message in $S_k$ which arrives sooner than it could in $S_k'$.  This occurs $d-u$ after time $t_{k-3}^*$, when the message delays changed.  The soonest $p_{k-1}$ can learn about the difference is when a message from $p_{k-2}$, sent after it detected the difference, could arrive.  But $t_{k-3}^* + (d-u) = t_{k-2}^*$, so any message $p_{k-2}$ sends to $p_{k-1}$ after this point has delay $d$.  Thus, the soonest $p_{k-1}$ can distinguish $S_k$ from $S_k'$ is $t_{k-2}^*+d = t_1 + (k-2)(d-u) + d$.  We argue that this is after $p_{k-1}$ generates its $Dequeue$ return value, which occurs no later than $t_1 + (k-1)s + |Dequeue|$.  We thus want to show that $t_1 + (k-2)(d-u) + d > t_1 + (k-1)s + |Dequeue|$.  Consider the cases for $s$:
\begin{itemize}
\item $s = 0$: The inequality holds iff $(k-2)(d-u) + d > |Dequeue|$, which is true because $d\geq u$ and $d > |Dequeue|$.
\item $s = Q - u$: The inequality holds if $(k-2)(d-u) + d > (k-1)(Q-u) + |Dequeue|$, or $(k-1)d > (k-1)Q - u + |Dequeue|$.

  Since $|Dequeue| < Q$, it is sufficient to show that $(k-1)d \geq kQ-u$, or $Q \leq \frac{(k-1)d + u}{k}$.

  To prove this final inequality, recall that $Q \leq \frac{3d+2u}{5}$ and that $k \geq 4$.  For all $k \geq 4$, $\frac{(k-1)d+u}{k} \geq \frac{3d+u}{4}$, so it is sufficient to show that $\frac{3d+2u}{5} \leq \frac{3d+u}{4}$.  This follows because $\frac{3d+2u}{5} = \left(\frac{3d+u}{4}\right)\left(\frac{4}{5}\right) + \frac{u}{5}$, and $\frac{u}{5} \leq \left(\frac{1}{5}\right)\left(\frac{3d+u}{4}\right)$, as that inequality reduces to $u \leq d$, which is true.
\end{itemize}

We conclude that $p_k$ cannot distinguish $S_k$ from $S_k'$ until after it generates a return value for its $Dequeue$ instance, so it must return the same value in both runs, which we previously proved was in the set $\{1,\dots,k-1\}$.

Now, by mathematical induction, when $k=n$, all $Dequeue$ instances in $S_n$ must return values from the set $\{1,\dots,n-1\}$, and we have the claim.
\end{proof}

\section{Contradiction}\label{sec:contradiction}

Let us quickly recap what we have shown so far.  First, we showed that there is a run $D_n$ with $n$ overlapping $Dequeue$ instances which must each return a different value.  Then, we (somewhat laboriously) showed that there is a run $S_n$ with $n$ overlapping $Dequeue$ instances in which at least two $Dequeue$ instances must return the same value.  Now, we want to show that these runs are indistinguishable, which leads to a contradiction, as processes must return the same values in indistinguishable runs, and no set of return values is simultaneously distinct and contains a repeated value.

\begin{theorem}
  There is no uniform, set-linearizable implementation of a multiplicity queue with $|Dequeue| < \min\left\{\frac{d}{2}+u, \frac{3d+2u}{5}\right\}$.
\end{theorem}

\begin{proof}
  Assume, in contradiction, that there is such an algorithm.  Then the conditions for Lemma~\ref{lem:distinctReturns} and Lemma~\ref{lem:sameReturns} are satisfied, so we know that $S_n$ and $D_n$ exist, where all $Dequeue$ instances in $S_n$ return values from $\{1,\dots, n-1\}$ and the $Dequeue$ instance at $p_i$ in $D_n$ returns $i+1$, for all $0 \leq i < n$.  Recall that $S_n$ requires that $n \geq N_0$, defined in Section~\ref{sec:sameValuesBaseCase} s.t. for all $n \geq N_0$, $|Dequeue| < \frac{d}{2} + {n-2}{n}u$.

  Note that $S_n$ and $D_n$ are nearly identical--they have the same initial sequence of $Enqueue$ instances at $p_0$, the same clock offsets ($c_0 = 0$, $c_i = \frac{i}{n}u, 1\leq i < n$), and the same $Dequeue$ invocations ($p_i$ invokes $Dequeue$ at time $t_1 + (i-1)s$).  The two runs also have nearly identical message delays, where if $0 \leq i < j < n$, messages from $p_j$ to $p_i$ have delay $d$ and those from $p_i$ to $p_j$ have delay $d-u$, except that in $S_n$, messages from $p_{n-2}$ to $p_{n-1}$ sent at or after time $t_{n-2}^*$ have delay $d$.  Thus, if we extend those message delays in $D_n$, we will have the same run.  We will argue that we will still have $D_n$'s behavior, which differs from $S_n$'s, in the same run, which is a contradiction.

  Suppose first that $u < d$.  Construct $D^*$ from $D_n$ by delaying all messages from $p_{n-2}$ to $p_{n-1}$ sent at or after $t_{n-2}^*$ by $d$.  We argue that no process can distinguish that it is in $D^*$ instead of $D_n$ before its $Dequeue$ instance returns.  The first point where any process could distinguish the two runs is when a message $p_{n-2}$ sends at $t_{n-2}^*$ does not arrive at $p_{n-1}$ at the same time in $D^*$ it would in $D_n$, because we extended its delay.  Thus, the first time a process can distinguish the two runs is $t_{n-2}^* + d-u = t_1 + (n-2)(d-u) + (d-u) = t_1 + (n-1)(d-u)$.  We argue that, for sufficiently large $n$, this is after $p_{n-1}$'s $Dequeue$ instance returns.  That happens at or before $t_1 + (n-1)s + Q$.  We thus want $t_1 + (n-1)(d-u) > t_1 + (n-1)s + Q$, which is true iff $(n-1)(d-u) > (n-1)s + Q$.  Consider the possible values of $s$ by cases:
  \begin{itemize}
  \item $s = 0$: We want to show that $(n-1)(d-u) > Q$.  This is true when $n > \frac{Q}{d-u} + 1$.  Since $d > u$, this is true for sufficiently large $n$. Let $N_1$ be such that for all $n \geq N_1, n > \frac{Q}{d-u} + 1$.
  \item $s = D-u$: We want to show that $(n-1)(d-u) > (n-1)(Q-u) + Q$.  This is true when $(n-1)d - (n-1)u > nQ - (n-1)u$, or $(n-1)d > nQ$.  If we solve for $n$, we have $n(d-Q) - d > 0$, or $n > \frac{d}{d-Q}$, since $d-Q>0$.  Again, this is true for sufficiently large $n$, so let $N_2$ be such that for all $n \geq N_2$, $n > \frac{d}{d-Q}$.
  \end{itemize}

  Thus, in runs with sufficiently large $n$ (at least $\max\{N_0,N_1,N_2\}$), $p_{n-1}$ cannot distinguish that it is in $D^*$, not $D_n$, until after its $Dequeue$ instance has returned.\footnote{We see here that our proof does not exactly apply only to uniform algorithms, but to any algorithm running on at least $\max\{N_0,N_1,N_2\}$ processes.  However, we state the result for uniform algorithms to get a result that applies to any size system, as we do not prove that non-uniform algorithms running on fewer processes cannot achieve higher performance.}  Similarly, no other process can distinguish the runs before its $Dequeue$ instance returns, as those returns occur by $t_1 + (i)s + Q \leq t_1 + (n-1)s + Q$ for $0\leq i < n$, so there is not time for $p_{n-1}$ to inform any other process of the discrepancy since by the time $p_{n-1}$ discovers it, all other processes' $Dequeue$ instances have already returned.

  Next, we have the case where $u = d$.  In this case, observe that $t_{n-2}^* = t_1 + (n-2)(d-u) = t_1$, so all messages from $p_{n-2}$ to $p_{n-1}$ starting at $t_1$ have delay $d$.  Thus, $p_{n-1}$ can distinguish the runs at $t_1 + (d-u) = t_1$, which is before its $Dequeue$ instance returns.

  Instead, we can use a reduction argument to disprove the existence of an algorithm performing better than our bound.  Choose a new message uncertainty $u' = \frac{d+|Dequeue|}{2}$, noting that this gives $0 < u' < d$.  Now, since our assumed algorithm correctly implements a multiplicity queue in a system with message delays in the range $[0,d]$ with $|Dequeue| < \min\left\{\frac{3d+2u}{5},\frac{d}{2}+u\right\} = d$, it must also correctly implement that multiplicity queue in a system with message delays $[d-u',d]$, since any run possible in that system is possible in the system where $d=u$ since the range of possible message delays is completely contained in $[d-u,d]$.  It thus implements multiplicity queues in a system with message uncertainty $u'$ with $|Dequeue| < d = 2u' - |Dequeue|$.  Then $|Dequeue| < u' < \min\left\{\frac{3d + 2u'}{5}, \frac{d}{2}+u'\right\}$ because $d > u'$.  But this contradicts the impossibility of such an algorithm as proved in the $u<d$ case above, so our assumed algorithm cannot exist.
\end{proof}
    
Finally, we note that our result is an improvement over the previously best-known bound from \cite{Talmage22}, with the added restriction to uniform algorithms.  This claim follows from elementary algebra, as $\frac{d+u}{2} = \frac{d}{2} + \frac{u}{2} < \frac{d}{2} + u$ and $\frac{d+u}{2} \leq \frac{2.5d + 2.5u}{5} \leq \frac{3d+2u}{5}$, since $u\leq d$.

\begin{corollary}
  Any uniform, set-linearizable implementation of a multiplicity queue must have $|Dequeue| \geq \frac{d+u}{2} \geq \min\left\{\frac{2d}{3},\frac{d+u}{2}\right\}$.
\end{corollary}

\section{Partial Tightness: Special Case Upper Bound}

While it may seem that the $\frac{d}{2} + u$ term in the lower bound is an artifact of our limited proof techniques for lower bounds, and future work may increase the bound to $\frac{3d+2u}{5}$ or better for all values of $u$, we here outline an algorithm for the special case where $u=0$ which matches the $\frac{d}{2} + u = \frac{d}{2}$ lower bound, beating $\frac{3d}{5}$.  This suggests $\frac{d}{2} + u$ may be somehow fundamental, despite not holding everywhere

The algorithm is event-driven, where each process can react to operation invocations, message receptions, and expiration of local timers it sets.  Because $u=0$, every message takes exactly $d$ time to arrive.  Thus, since the algorithm broadcasts every message, when any process receives a message, it knows all other processes receive the same message at the same time.  Further, since there is no uncertainty, the maximum clock skew is $(1-1/n)0 = 0$, so every process' local clock (read by the function $localClock()$) is equal to real time.  We thus let every operation instance take $d/2$ time.  By the message delay and operation instance duration, a process learns about an instance at another process before it returns to an instance at itself if and only if that remote instance returned before the local one's invocation, so applying remote operations to the local copy of the structure immediately upon receipt and choosing $Dequeue$ return values $d/2$ after invocation together keep the local copies synchronized and choose correct values.

\begin{algorithm}
  \caption{Set-linearizable implementation of a multiplicity queue with $u=0$. Code for each $p_i$}\label{alg:zeroU}
  \begin{algorithmic}[1]
    \algtext*{EndFunction}
    \algtext*{EndIf}
    \Statex \textbf{Initially:} $localQueue$ is an empty FIFO queue, $mostRecentDequeue = 0$
    \algrenewcommand\algorithmicfunction{\textbf{HandleInvocation}}
    \Function{Enqueue}{$arg$}
    \State send $\langle enq, arg\rangle$ to all other processes
    \State $setTimer(d/2, \langle enq,arg,\langle localClock(), i\rangle, return\rangle)$\label{line:enqSetTimer}
    \EndFunction

    \Function{Dequeue}{}
    \State send $\langle deq, ts=\langle localClock(), i\rangle\rangle$ to all other processes\label{line:sendDeq}
    \State $setTimer(d/2, \langle deq, ts\rangle)$\label{line:deqSetTimer}
    \EndFunction

    \algrenewcommand\algorithmicfunction{\textbf{HandleTimer}}
    \Function{Expire}{$\langle enq, arg, ts, return\rangle$}
    \State Generate $Enqueue$ response to user
    \State $setTimer(d/2, \langle enq, arg, apply\rangle)$
    \EndFunction

    \Function{Expire}{$\langle enq, arg, apply\rangle$}
    \State $localQueue.enqueue(arg)$
    \EndFunction

    \Function{Expire}{$\langle deq, \langle clockVal, i\rangle\rangle$}
    \State Generate $Dequeue$ response to user with return value $localQueue.dequeue()$\label{line:deqReturn}
    \State $mostRecentDequeue = clockVal$
    \EndFunction

    \algrenewcommand\algorithmicfunction{\textbf{HandleReceive}}
    \Function{$\langle enq, arg\rangle$}{}
    \State $localQueue.enqueue(arg)$
    \EndFunction

    \Function{$\langle deq, \langle clockVal, j\rangle\rangle$}{}
      \If {$clockVal > mostRecentDequeue + d/2$}\label{line:deqSpacing}
        \State $localQueue.dequeue()$\label{line:localDeq}
        \State $mostRecentDequeue = clockVal$
      \EndIf
    \EndFunction 
  \end{algorithmic}
\end{algorithm}

Let $R$ be an arbitrary run of Algorithm~\ref{alg:zeroU}.  Observe that every invocation in $R$ either has a matching response, $d/2$ after invocation.  We define a set-linearization of $R$, prove that it respects real time order and that it is a legal sequence of sets, and we have the correctness of the algorithm.

\begin{construction}\label{constr}
  Place each $Enqueue$ instance in a singleton set and define the set's timestamp as the pair of the invoking process' local clock read on line~\ref{line:enqSetTimer} plus $d/2$ and the invoking process' id.  For each $Dequeue$ return value $x$, place all $Dequeue$ instances which return $x$ in a set, and define the set's timestamp as the smallest timestamp of any instance in the set, where a $Dequeue$ instance's timestamp is the pair of the local clock read in line~\ref{line:sendDeq} and the invoking process' id, with the id breaking ties between clock values.  Let $\pi$ be the sequence of these sets ordered by increasing timestamps (break ties by process id).
\end{construction}

\begin{lemma}\label{lem:realTimeOrder}
  $\pi$ respects the order of non-overlapping operation instances.
\end{lemma}

\begin{proof}
  Let $op_1$ and $op_2$ be any two non-overlapping operation instances, with $op_1$ invoked at $p_i$ and returning before $op_2$'s invocation at $p_j$.  Since local clocks are exactly real time, and all instances have duration $d/2$, then $op_1$'s timestamp will be more than $d/2$ smaller than $op_2$'s.  Thus, the only way that $op_1$ would not strictly precede $op_2$ in $\pi$ is if they were in the same set, which could happen if they are both $Dequeue$ instances which returned the same value $x$.  But in that case, since $op_1$ returned before $op_2$'s invocation and each of $op_1$ and $op_2$ took $d/2$ time between invocation and response, then $p_j$ would receive the message $p_i$ sent on line~\ref{line:sendDeq} at $op_1$'s invocation before $op_2$ returns.  This should have removed $x$ from $p_j$'s local copy of the queue, unless there were another element preceding $x$ in $p_j$'s local queue when $op_2$ returned.  By the FIFO ordering of the multiplicity queue, this can only happen if there is a $Dequeue$ instance which $p_i$ applied before $op_1$ returned but $p_j$ did not apply before $op_2$ returned.  Any $Dequeue$ instance which $p_i$ has applied before $op_1$ returns was either delivered to $p_j$ at the same time as to $p_i$, and thus applied to $p_j$'s local copy or invoked at $p_i$ before $op_1$, but then by the time $op_1$ returns, by the fact that every $Dequeue$ returns $d/2$ time after invocation, $p_j$ would also receive and apply that $Dequeue$ instance before $op_2$'s invocation.  Thus, there cannot be an element in $p_j$'s local queue preceding $x$ when it applies $op_1$, and $op_2$ cannot return $x$.
\end{proof}

\begin{lemma}\label{lem:legalSequence}
  $\pi$ is legal by the specification of a multiplicity queue.
\end{lemma}

\begin{proof}
  We proceed by induction on $\sigma$, a prefix of $\pi$.  If $\sigma$ is empty, then it is legal, as the empty sequence is always legal.

  Suppose that $\sigma = \rho \cdot S$, where $S$ is a set of operation instances.  Assume that $\rho$ is legal.  We will show that $\sigma$ is also legal by cases on $S$.

  If $S = Enqueue(x)$, then $\sigma$ is necessarily legal, as $Enqueue$ does not return a value, so cannot be illegal.

  If $S$ is a set of $Dequeue$ instances returning $x$, then we need to argue that the algorithm chose $x$ correctly.  Each invoking process chose the oldest value in its local copy of the queue as a return value, in line~\ref{line:deqReturn}, so we merely need to argue that the local copy of the queue contains the elements enqueued and not dequeued in $\rho$, in order.  Consider the $Dequeue$ instance in $S$ with the smallest timestamp, and call it $d$ and its invoking process $p_i$.  When $p_i$ executes line~\ref{line:deqReturn} to generate $d$'s return, it will have received every $Enqueue$ invocation in $\rho$, as those were invoked at least $d/2$ before than this $Dequeue$, and added them to its local queue.  The order of $Enqueue$ instances in $\rho$ matches their timestamp order, which is the order in which they are locally applied, since every process adds each $Enqueue$ argument $d$ time after its invocation.  When any other process $p_j$ which has a $Dequeue$ instance return the same value as $d$ executes line~\ref{line:deqReturn} for that instance, it will have locally applied all $Enqueue$ instances $p_i$ has, and possible more.  But any additional $Enqueue$ instances will have larger timestamps, and thus follow $Enqueue(x)$ in $\pi$, so would not be the correct return value for this $Dequeue$ instance.

  Thus, each process chooses $x$ as the oldest-enqueued value in $\rho$ which it has not already removed for another $Dequeue$ instance.  Such an instance must be in $\rho$, as another $Dequeue$ instance at the same process would have a smaller timestamp and one at another process would not remove a value from the local queue until $d$ after its invocation, which means it would have a smaller timestamp than this $Dequeue$ instance which returns $x$.

  Further, each process only removes values from its local queue when there is a $Dequeue$ instance returning it.  Suppose this were not so.  Then some process $p_k$ must have received a $Dequeue$ instance which returned $y$ and executed line~\ref{line:localDeq} when it had already removed $y$ from its local queue.  But $p_k$ could only remove $y$ when it either returned $y$ to its own $Dequeue$ instance or received a message about another $Dequeue$ instance.  But either of those cases would update $mostRecentDequeue$, so the check on line~\ref{line:deqSpacing} means that the two $Dequeue$ instances which returned $y$ had timestamps more than $d/2$ apart, which implies they were not concurrent, so they could not have returned the same value as that would imply they are in the same set in $\pi$, which is not possible by Lemma~\ref{lem:realTimeOrder}.

  Finally, there cannot be a $Dequeue$ instance returning $x$ in $\rho$, as all instances returning $x$ are in the set $S$.  Thus, $x$ is the argument of the first $Enqueue$ instance in $\rho$ which is not returned by a $Dequeue$ instance in $\rho$.
\end{proof}

\begin{theorem}
  If $u = 0$, Algorithm~\ref{alg:zeroU} is a uniform, set-linearizable implementation of a multiplicity queue with $|Dequeue| = d/2$.
\end{theorem}

\begin{proof}
  By Lemma~\ref{lem:realTimeOrder}, the sequence $\pi$ we defined in Construction~\ref{constr} respects the real-time order of non-overlapping instances.  Lemma~\ref{lem:legalSequence} proves that $\pi$ is legal, so it is a legal set-linearization, proving by construction that Algorithm~\ref{alg:zeroU} is a set-linearizable implementation of a multiplicity queue.  By lines~\ref{line:deqSetTimer} and \ref{line:deqReturn}, every $Dequeue$ instance returns $d/2$ time after invocation, so $|Dequeue| = d/2$.  Finally, the code for Algorithm~\ref{alg:zeroU} does not depend on $n$, so it is a uniform algorithm.
\end{proof}
Since this matches our lower bound of $|Dequeue| \geq \min\left\{\frac{3d+2u}{5},\frac{d}{2} + u\right\} = \frac{d}{2}$ when $u=0$, this algorithm is optimal and proves the bound is tight in this case.

\section{Conclusion}

We developed a new combination of shifting and other indistinguishability arguments to prove a larger lower bound of $|Dequeue| \geq \min\left\{\frac{3d+2u}{5},\frac{d}{2} + u\right\}$ in uniform multiplicity queue implementations.  This both improves the state of the art and suggests ways to improve the bound further.  For example, strengthening the base case for Lemma~\ref{lem:sameReturns} in Section~\ref{sec:sameValuesBaseCase} should improve the $\frac{3d+2u}{5}$ portion of the lower bound.  We hypothesize that this may increase to approach a limit of $|Dequeue| \geq d$ for all non-zero values of $u$, which seems an intuitive value.  If that is true, our tightness result that $|Dequeue| = d/2$ is possible when $u=0$ is more interesting, as it suggests the bounds may be discontinuous.  We continue exploring these bounds to understand multiplicity queues, and then use that understanding to design and understand other data type relaxations.  

\section{References}

\bibliographystyle{plain}
\bibliography{../../refs.bib}

\begin{thebibliography}{10}

\bibitem{AfekKorlandYanovsky10}
Yehuda Afek, Guy Korland, and Eitan Yanovsky.
\newblock Quasi-linearizability: Relaxed consistency for improved concurrency.
\newblock In Chenyang Lu, Toshimitsu Masuzawa, and Mohamed Mosbah, editors,
  {\em Principles of Distributed Systems - 14th International Conference,
  {OPODIS} 2010, Tozeur, Tunisia, December 14-17, 2010. Proceedings}, volume
  6490 of {\em Lecture Notes in Computer Science}, pages 395--410. Springer,
  2010.

\bibitem{AttiyaBar-NoyDolev95}
Hagit Attiya, Amotz Bar{-}Noy, and Danny Dolev.
\newblock Sharing memory robustly in message-passing systems.
\newblock {\em J. {ACM}}, 42(1):124--142, 1995.

\bibitem{CastanedaPina21}
Armando Casta{\~{n}}eda and Miguel Pi{\~{n}}a.
\newblock Fully read/write fence-free work-stealing with multiplicity.
\newblock In Seth Gilbert, editor, {\em 35th International Symposium on
  Distributed Computing, {DISC} 2021, October 4-8, 2021, Freiburg, Germany
  (Virtual Conference)}, volume 209 of {\em LIPIcs}, pages 16:1--16:20. Schloss
  Dagstuhl - Leibniz-Zentrum f{\"{u}}r Informatik, 2021.

\bibitem{CastanedaRajsbaumRaynal20}
Armando Casta{\~{n}}eda, Sergio Rajsbaum, and Michel Raynal.
\newblock Relaxed queues and stacks from read/write operations.
\newblock In Quentin Bramas, Rotem Oshman, and Paolo Romano, editors, {\em 24th
  International Conference on Principles of Distributed Systems, {OPODIS} 2020,
  December 14-16, 2020, Strasbourg, France (Virtual Conference)}, volume 184 of
  {\em LIPIcs}, pages 13:1--13:19. Schloss Dagstuhl - Leibniz-Zentrum f{\"{u}}r
  Informatik, 2020.

\bibitem{HenzingerKirschPayerSezginSokolova13}
Thomas~A. Henzinger, Christoph~M. Kirsch, Hannes Payer, Ali Sezgin, and Ana
  Sokolova.
\newblock Quantitative relaxation of concurrent data structures.
\newblock In Roberto Giacobazzi and Radhia Cousot, editors, {\em The 40th
  Annual {ACM} {SIGPLAN-SIGACT} Symposium on Principles of Programming
  Languages, {POPL} '13, Rome, Italy - January 23 - 25, 2013}, pages 317--328.
  {ACM}, 2013.

\bibitem{Herlihy91}
Maurice Herlihy.
\newblock Wait-free synchronization.
\newblock {\em {ACM} Trans. Program. Lang. Syst.}, 13(1):124--149, 1991.

\bibitem{JohnenKhattabiMilani22}
Colette Johnen, Adnane Khattabi, and Alessia Milani.
\newblock Efficient wait-free queue algorithms with multiple enqueuers and
  multiple dequeuers.
\newblock In Eshcar Hillel, Roberto Palmieri, and Etienne Rivi{\`{e}}re,
  editors, {\em 26th International Conference on Principles of Distributed
  Systems, {OPODIS} 2022, December 13-15, 2022, Brussels, Belgium}, volume 253
  of {\em LIPIcs}, pages 4:1--4:19. Schloss Dagstuhl - Leibniz-Zentrum
  f{\"{u}}r Informatik, 2022.

\bibitem{KhanchandaniWattenhofer18}
Pankaj Khanchandani and Roger Wattenhofer.
\newblock On the importance of synchronization primitives with low consensus
  numbers.
\newblock In Paolo Bellavista and Vijay~K. Garg, editors, {\em Proceedings of
  the 19th International Conference on Distributed Computing and Networking,
  {ICDCN} 2018, Varanasi, India, January 4-7, 2018}, pages 18:1--18:10. {ACM},
  2018.

\bibitem{Kosa99}
Martha~J. Kosa.
\newblock Time bounds for strong and hybrid consistency for arbitrary abstract
  data types.
\newblock {\em Chic. J. Theor. Comput. Sci.}, 1999, 1999.

\bibitem{LundeliusLynch84}
Jennifer Lundelius and Nancy~A. Lynch.
\newblock An upper and lower bound for clock synchronization.
\newblock {\em Information and Control}, 62(2/3):190--204, 1984.

\bibitem{Neiger94}
Gil Neiger.
\newblock Set-linearizability.
\newblock In James~H. Anderson, David Peleg, and Elizabeth Borowsky, editors,
  {\em Proceedings of the Thirteenth Annual {ACM} Symposium on Principles of
  Distributed Computing, Los Angeles, California, USA, August 14-17, 1994},
  page 396. {ACM}, 1994.

\bibitem{ShavitTaubenfeld16}
Nir Shavit and Gadi Taubenfeld.
\newblock The computability of relaxed data structures: queues and stacks as
  examples.
\newblock {\em Distributed Comput.}, 29(5):395--407, 2016.

\bibitem{Talmage22}
Edward Talmage.
\newblock Lower bounds on message passing implementations of
  multiplicity-relaxed queues and stacks.
\newblock In Merav Parter, editor, {\em Structural Information and
  Communication Complexity - 29th International Colloquium, {SIROCCO} 2022,
  Paderborn, Germany, June 27-29, 2022, Proceedings}, volume 13298 of {\em
  Lecture Notes in Computer Science}, pages 253--264. Springer, 2022.

\bibitem{TalmageWelch14}
Edward Talmage and Jennifer~L. Welch.
\newblock Improving average performance by relaxing distributed data
  structures.
\newblock In Fabian Kuhn, editor, {\em Distributed Computing - 28th
  International Symposium, {DISC} 2014, Austin, TX, USA, October 12-15, 2014.
  Proceedings}, volume 8784 of {\em Lecture Notes in Computer Science}, pages
  421--438. Springer, 2014.

\bibitem{TalmageWelch18}
Edward Talmage and Jennifer~L. Welch.
\newblock Relaxed data types as consistency conditions.
\newblock {\em Algorithms}, 11(5):61, 2018.

\bibitem{TalmageWelch19}
Edward Talmage and Jennifer~L. Welch.
\newblock Anomalies and similarities among consensus numbers of
  variously-relaxed queues.
\newblock {\em Computing}, 101(9):1349--1368, 2019.

\bibitem{WangTalmageLeeWelch18}
Jiaqi Wang, Edward Talmage, Hyunyoung Lee, and Jennifer~L. Welch.
\newblock Improved time bounds for linearizable implementations of abstract
  data types.
\newblock {\em Inf. Comput.}, 263:1--30, 2018.

\end{thebibliography}


\end{document}